# The Role of Schwartz Measures in Human Tri-Color Vision

M. L. Sloan
Austin Research Associates

## Abstract

The human tri-color vision process may be characterized as follows:

1. A requirement of three scalar quantities to fully define a color (for example, intensity, hue, and purity), with

2. These scalar measures linear in the intensity of the incident light, allowing in general any specific color to be duplicated by an additive mixture of light from three standardized (basis) colors,

3. The exception being that the spectral colors are unique, in that they cannot be duplicated by any positive mixture of other colors.

These characteristics strongly suggest that human color vision makes use of Schwartz measures in processing color data. This hypothesis is subject to test. In this brief paper, the results of this hypothesis are shown to be in good agreement with measured data.

## **Overview of Schwartz Measures**

Schwartz measures arise in consideration of the non-negative character of integrals of the form[1]

$$I(K) = \int S(\lambda) W(\lambda) [F(\lambda) - K]^2 d\lambda \geq 0 \qquad (1)$$

and play a role in such diverse areas as thermodynamics of gases, quantum mechanics, and dollar-cost-averaging. Here $S(\lambda)$ is a non-negative distribution or "intensity" function of the parameter $\lambda$; $W(\lambda)$, a non-negative weighting function; and $F(\lambda)$, a "moment" function of $\lambda$ generally monotonic in $\lambda$. The integration is over a specified range of $\lambda$; the quantity $K$ is a constant.

An important feature of Eq. (1) is that the integral is zero if and only if:

1) The distribution function $S(\lambda)$ is a delta function in $\lambda$:

$$S(\lambda) = \delta(\lambda - \lambda_o)$$

---

[1] An equivalent expression often more suited for analysis is
$I(K) = \int S(\lambda) [f(\lambda) - K g(\lambda)]^2 d\lambda \geq 0.$

and   2)  $K = F(\lambda_o)$.

Explicitly carrying out the expansion of the quadratic in Eq. (1), we have:

$$I(K) = K^2\, [1] + 2\, K\, [F] + [FF] \geq 0 \tag{2}$$

where

$$[1] \equiv \int S(\lambda)\, W(\lambda)\, d\lambda \qquad \geq 0 \quad \text{by construction}$$

$$[F] \equiv \int S(\lambda)\, W(\lambda)\, F(\lambda)\, d\lambda$$

$$[FF] \equiv \int S(\lambda)\, W(\lambda)\, F(\lambda)^2\, d\lambda \qquad \geq 0 \quad \text{by construction}$$

are the Schwartz measures: More precisely, the zero, first, and second moment of the distribution function $S(\lambda)$ with respect to $F(\lambda)$.

Since $I(K)$ is a non-negative quadratic in $K$, it has a minimum. Evaluating Eq. (2) at that minimum K, one obtains the classic Schwartz inequality

$$[FF]\,[1] \geq [F]^2 \quad \text{for all } \lambda. \tag{3}$$

As has already been pointed out, the equality in Eq. (3) holds if and only if the distribution function $S(\lambda)$ is a delta function. For all other distribution functions,

$$[FF]\,[1] \;>\; [F]^2\,.$$

If we now identify the distribution function $S(\lambda)$ with the intensity of a particular light source as a function of the wavelength $\lambda$, we can make the following observation:

Were Schwartz measures and the Schwartz inequality of Eq. (3) to play a role in human color vision, it would be necessary that

1. The human color process be characterized by three scalar measures (in this case, [1], [F], [FF] or linear combinations thereof);

2. These measures would moreover be linear in the light intensity $S(\lambda)$;

and, finally,

3. The topology of color space would be such that spectrally pure colors (i.e. delta function sources in $\lambda$) are unique in that it is only for such sources that $[FF][1] = [F]^2$. For all other, non-pure sources, $[FF][1] > [F]^2$.

The fact that human color vision appears to qualitatively conform exactly to these requirements strongly suggests that the color vision process may indeed make use of such measures.[2] This hypothesis is subject to quantitative tests, as now demonstrated.

---

[2] Under such a development, the quantity $P \equiv [F]^2 / ([1][FF])$ might serve as a good surrogate for color purity, since $0 \leq P \leq 1$, with $P = 1$ only for spectral colors. It is also easy to show that for a color C (of purity P) resulting from a positive mixture of colors $C_0$ (purity $P_0$) and $C_1$ (Purity $P_1$), the resulting P is always less than the maximum of ($P_0$, $P_1$). In other words, the purity of C is always less than the maximum purity of the colors from which it is derived.

Moreover, the quantity $H = [F]/[1]$ might serve as a good surrogate for hue. Since the perceived hue is different for each spectral color, the quantity $H$, and hence the function $F(\lambda)$, would need to be monotonic in $\lambda$ over the visible spectrum.

## Tests of the Schwartz Measure Hypothesis

Test 1. Agreement with Tristimulus Data

The Tristimulus Values X, Y, Z are often used as three parameters specifying a color. These values are themselves measures of the intensity function folded with various weighting functions, namely the Spectral Tristimulus Values as given, for example, in the C.E.I. 1931 Standard [Ref. 1].

If the color vision process utilizes Schwartz measures [1], [F], [FF] in processing color information, then it must be true that the three Schwartz measures form an equally valid basis for defining color and, therefore, must be expressible as linear combinations of the Tristimulus values, and vice versa. Specifically, then, there must exist three three-dimensional constant vectors, **A**, **B**, **C**, such that

$$[1] = A_1 X + A_2 Y + A_3 Z$$

$$[F] = B_1 X + B_2 Y + B_3 Z \tag{4}$$

$$[FF] = C_1 X + C_2 Y + C_3 Z$$

Again, for non-spectral colors, $[FF][1] > [F]^2$. However, for the pure spectral colors, $[FF][1] = [F]^2$, which requires that the spectral Tristimulus values obey:

$$(A_1 X + A_2 Y + A_3 Z)(C_1 X + C_2 Y + C_3 Z) = (B_1 X + B_2 Y + B_3 Z)^2 \tag{5}$$

Carrying out the algebra, we see that the Tristimulus spectral locus must be defined by a 5 parameter quadratic form (a conic section):

$$r X^2 + Y^2 + s Z^2 + t XZ + u XY + v YZ = 0 \tag{6}$$

where
$$\begin{aligned}
A_1 C_1 - B_1^2 &= (A_2 C_2 - B_2^2) \, r \\
A_3 C_3 - B_3^2 &= (A_2 C_2 - B_2^2) \, s \\
A_1 C_3 + A_3 C_1 - 2 B_1 B_3 &= (A_2 C_2 - B_2^2) \, t \\
A_1 C_2 + A_2 C_1 - 2 B_1 B_2 &= (A_2 C_2 - B_2^2) \, u \\
A_2 C_3 + A_3 C_2 - 2 B_2 B_3 &= (A_2 C_2 - B_2^2) \, v
\end{aligned} \tag{7}$$

Demonstration of a set of constants r, s, t, u, and v such that Eq. (6) properly describes the measured spectral locus of color space, while at the same time providing real valued Eq. (7) solutions for **A**, **B**, and **C** provides a compelling test of this hypothesis.

Such a set of constants can in fact be found. [3]   Specifically, the Tristimulus spectral locus is found to be given by:

$$1.04592 \, X^2 - 2.72974 \, XY + Y^2 - 29.4858 \, XZ - 4.25065 \, YZ + 4.316942 \, Z^2 = 0 \qquad (8)$$

with resulting solutions for **A**, **B**, and **C** and the transformation between the Tristimulus data and the Schwartz measures given by:

$$
\begin{aligned}
{[1]} &= 1.84266 \, X + 1.19349 \, Y + 1.19349 \, Z \\
{[F]} &= 1.39833 \, X + 0.326642 \, Y - 0.673358 \, Z \\
{[FF]} &= 1.00181 \, X + 0.00181 \, Y + 0.00181 \, Z
\end{aligned}
\qquad (9)
$$

That Eq. (8) provides a good match to the known locus is best illustrated using chromaticity coordinates x, y. Since Eq. (7) is second order homogeneous in X, Y, Z, the equation can be transformed to the usual set of chromaticity coordinates:

$$x = X / (X + Y + Z)$$

$$y = Y / (X + Y + Z)$$

---

[3]   The reader may note that although there are 9 constants relating the tristimulus values to the Schwartz measures (Eq. 4), the Eq. (6) fit to the spectral locus involves only 5 numerical parameters.

The reason for this is that there is a four parameter subset of linear transformations amongst the Schwartz measures that automatically preserve the Schwartz inequality. In particular, for constants  a, b, c, d , with  $(a\,d - b\,c) \neq 0$, the following transformations:

$$
\begin{aligned}
{[1]}' &= c^2 \, [1] + 2\,c\,d \, [F] + d^2 \, [FF] \\
{[F]}' &= a\,c \, [1] + (a\,d + b\,c) \, [F] + b\,d \, [FF] \\
{[FF]}' &= a^2 \, [1] + 2\,a\,b \, [F] + b^2 \, [FF]
\end{aligned}
$$

are both necessary and sufficient to preserve the Schwartz inequality:

$$[FF]'\,[1]' - [F]'^{\,2} = (a\,d - b\,c)^2 \, ([FF]\,[1] - [F]^2) \geq 0$$

Thus the Eq. (9)  **A**, **B**, and **C**  solutions are only 1 of a 4-fold family of solutions. This freedom is useful in tailoring Schwartz measures to specific requirements, such as crafting a hue function H, as defined in Footnote 2,  that is monotonic in wavelength.

and the resulting locus from Eq. (7):

$$3.64298\, x^2 + 4.14348\, xy + y^2 - 3.98472\, x - 1.34672\, y + 0.45125 = 0 \qquad (10)$$

plotted against the chromaticity coordinates of the 1931 CEI standard. Results of this comparison are shown in Figure 1.

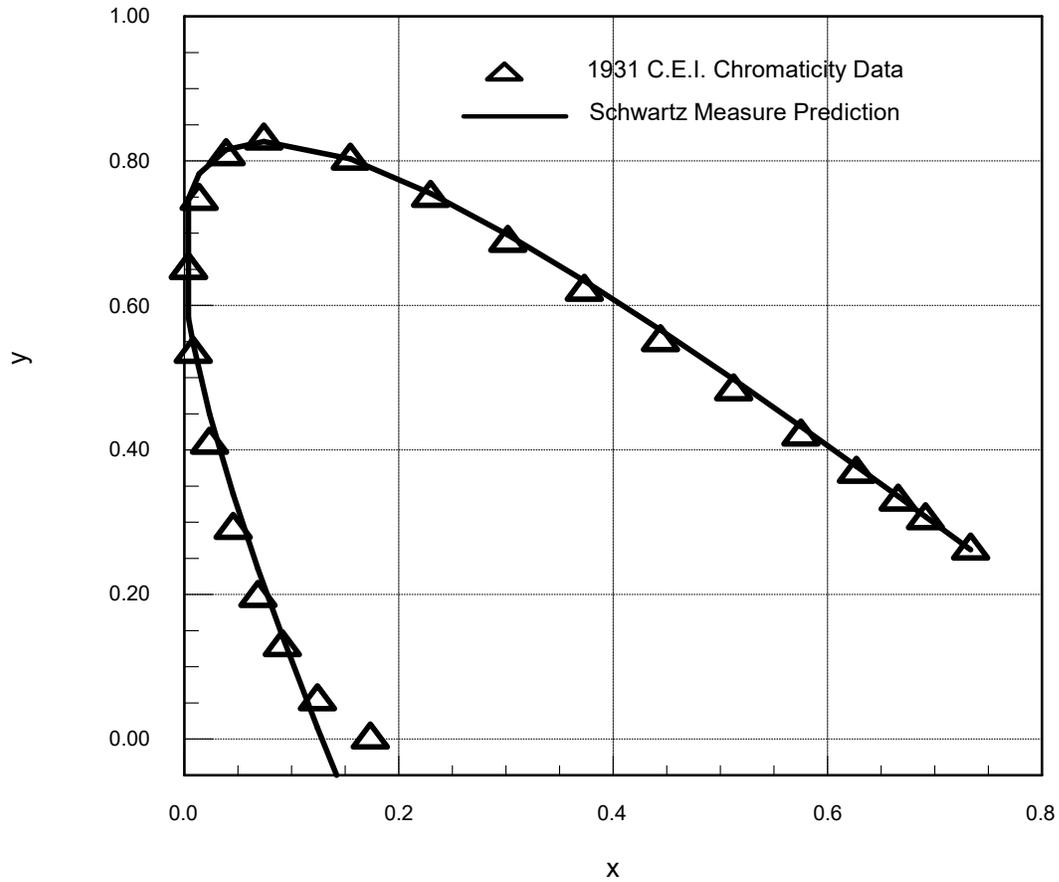

**Figure 1.** Comparison of spectral locus predicted by Schwartz measure hypothesis with 1931 C.E.I. Chromaticity Data.

Excellent agreement is noted, with the exception of the extreme end of the short wavelength blue region (400 nm).

Test 2.  Agreement of Schwartz Measures with Human Color Pigment Responses

It is generally recognized that the normal human color process involves the stimulation of three distinct wavelength-sensitive photoreceptor pigments [References 2,3].  Data of measurements by Boynton [Ref 2] showing the spectral responsivities of these three photoreceptors were presented in Figure 2 of an article by Robertson in Physics Today [Ref. 4].  Measurements from this figure were used to construct the following table of photoreceptivities, denoted R , G , B .[4]

**Table 1.**  Spectral Responses of Red (R), Green (G), and Blue (B) Photoreceptor Pigments

| λ [nm] | R | G | B |
|---|---|---|---|
| 400 | 0.000 | 0.000 | 0.121 |
| 420 | 0.008 | 0.010 | 0.649 |
| 440 | 0.016 | 0.031 | 1.000 |
| 470 | 0.060 | 0.100 | 0.660 |
| 480 | 0.100 | 0.160 | 0.460 |
| 485 | 0.130 | 0.200 | 0.360 |
| 490 | 0.163 | 0.263 | 0.274 |
| 495 | 0.200 | 0.316 | 0.221 |
| 500 | 0.263 | 0.395 | 0.169 |
| 505 | 0.326 | 0.495 | 0.132 |
| 510 | 0.411 | 0.600 | 0.105 |
| 515 | 0.511 | 0.705 | 0.079 |
| 520 | 0.611 | 0.816 | 0.060 |
| 530 | 0.774 | 0.942 | 0.032 |
| 540 | 0.884 | 0.986 | 0.016 |
| 550 | 0.947 | 0.984 | 0.005 |
| 560 | 0.989 | 0.921 | 0.000 |
| 570 | 0.999 | 0.814 | 0.000 |
| 580 | 0.968 | 0.653 | 0.000 |
| 590 | 0.895 | 0.484 | 0.000 |
| 600 | 0.789 | 0.316 | 0.000 |
| 610 | 0.663 | 0.200 | 0.000 |
| 620 | 0.526 | 0.113 | 0.000 |
| 680 | 0.024 | 0.000 | 0.000 |

---

[4] Measurements were at best only two digit accuracy.  Scaling the data to unit maximum response resulted in the three digit representation depicted here.

If these photoreceptivities form the physiological basis for human color vision and if such processes do indeed correspond to Schwartz measurements, then there must also exist linear combinations of R, G, B corresponding to the three Schwartz measures and these combinations must obey the Schwartz requirement $[FF][1] = [F]^2$ as a function of $\lambda$.

Again, such a combination can be found:

$$[1] = 0.729\,R - 0.271\,G + 0.729\,B$$

$$[F] = 0.500\,R - 0.327\,G - 0.500\,B \quad (11)$$

$$[FF] = 0.344\,R - 0.282\,G + 0.344\,B$$

and the fact that the R, G, B photoreceptivities obey the Schwartz requirement:

$$(0.729\,R - 0.271\,G + 0.729\,B)(0.344\,R - 0.281\,G + 0.344\,B)$$
$$= (0.500\,R - 0.327\,G - 0.500\,B)^2 \quad (12)$$

is amply demonstrated by the data of Figure 2. As in Figure 1, chromaticity-type coordinates

$$r = R / (R + G + B)$$

$$g = G / (R + G + B)$$

may defined and the solutions in r,g space of Eq. (12) plotted against the spectral response data of Table 1.

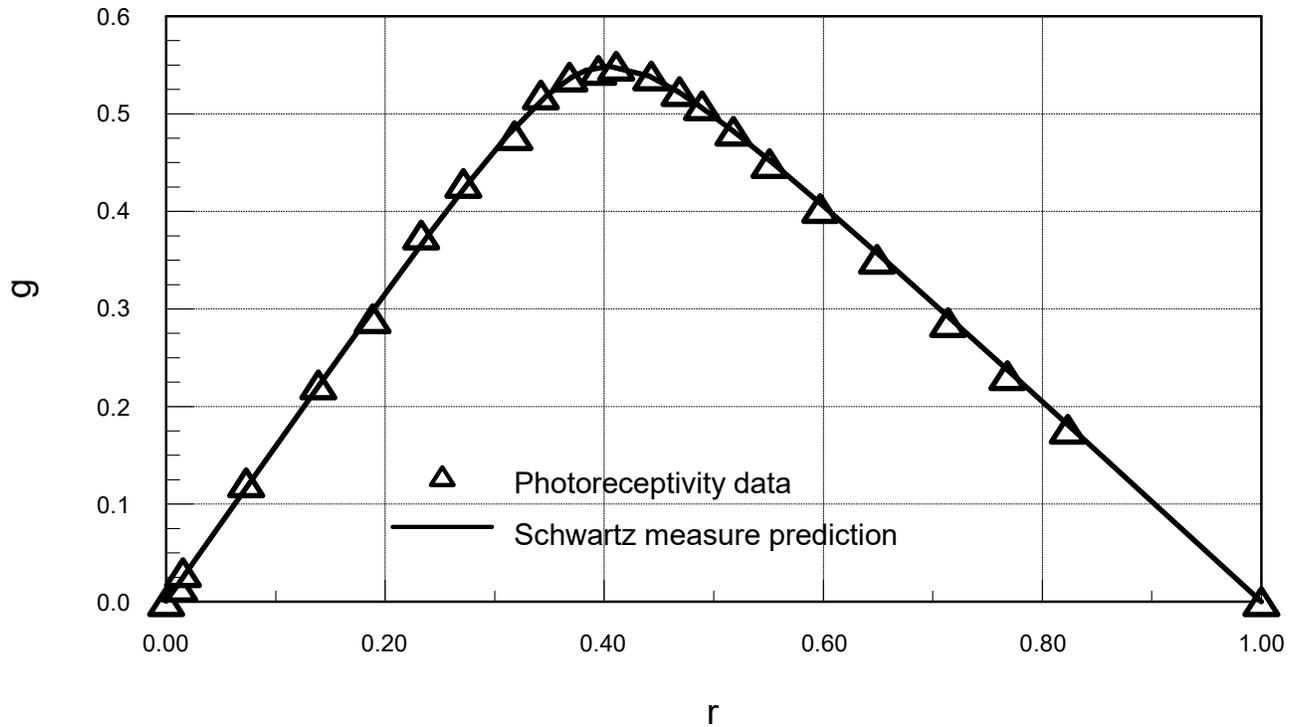

**Figure 2.**   Comparison of human photoreceptor pigment data with that predicted by the Schwartz measure hypothesis

Again the comparison is excellent.[5]

**Conclusions**

Both the standard spectral tristimulus data as well as measured human photoreceptor responsivities strongly support the hypothesis that the human color vision process makes use of Schwartz measures.

It would be interesting to determine whether these specific Schwartz measures, or transformations thereof as discussed in footnote 3, could provide the basis for a chromaticity metric.  But such considerations, as well as other interesting avenues of investigation, are beyond the scope of this brief paper.

---

[5] It seems to this author to be a profound statement about the physiology of the human vision process that the three color receptor pigment responsivities of the human eye cannot be arbitrary; but rather are intimately linked by the quadratic relationship of Eq. (12), the Schwartz requirement.